\documentclass[conference,letterpaper]{IEEEtran}

\usepackage[utf8]{inputenc} 
\usepackage[T1]{fontenc}
\usepackage{url}
\usepackage{ifthen}
\usepackage{cite}
\usepackage[cmex10]{amsmath} 

\usepackage{xcolor}
\usepackage{tabularx,ragged2e,booktabs}
\usepackage{mdwlist}
\usepackage{times}
\usepackage{float}
\usepackage{afterpage}

\usepackage{amstext}
\usepackage{amssymb,bm}
\usepackage{latexsym}
\usepackage{color}

\usepackage{pstricks}
\usepackage{enumerate}
\usepackage{mathtools}
\DeclarePairedDelimiter\ceil{\lceil}{\rceil}
\DeclarePairedDelimiter\floor{\lfloor}{\rfloor}

\newtheorem{thm}{Theorem}
\newtheorem{thm*}{Theorem}
\newtheorem{lemma}{Lemma}

\newtheorem{remark}{Remark}
\newtheorem{example}{Example}

\newcommand{\mais}{\text{MAIS}}
\newcommand{\smin}{s_{\min}}
\newcommand{\smax}{s_{\max}}

\begin{document}
\title{An Information Theoretic Converse for the ``Consecutive Complete--$S$'' PICOD Problem} 


\author{%
\IEEEauthorblockN{%
Tang Liu and Daniela Tuninetti\\%
University of Illinois at Chicago, Chicago, IL 60607 USA, 
Email: {\tt tliu44, danielat@uic.edu}\\%
}%
}
\maketitle

\begin{abstract}
Pliable Index CODing (PICOD) is a variant of the Index Coding (IC) problem in which a user is satisfied whenever it can successfully decode any one message that is not in its side information set, as opposed to a fixed pre-determined message.
The complete--$S$ PICOD with $m$ messages, for $S\subseteq[0:m-1]$,
has $n = \sum_{s\in S} \binom{m}{s}$ users with distinct side information sets.
Past work on PICOD provided tight converse results when either the sender is constrained to use linear codes, or for some special classes of complete--$S$ PICOD. 
This paper provides a tight information theoretic converse result (i.e., no restriction to linear codes) for the so-called ``consecutive complete--$S$'' PICOD, where the set $S$ satisfies $S=[\smin :\smax]$ for some $0\leq \smin\leq \smax \leq m-1$. 
This result extends existing converse results and shows that linear codes have the smallest possible code length given by $\min(m-\smin,1+\smax)$. The central contribution is a novel proof technique rooted in combinatorics.
The main idea is to consider all the messages a user can eventually successfully decode, in addition to its own desired message. 
This allows us to circumvent the necessity of essentially considering all possible assignments of desired messages for the users. The keystone of the proof is to show that, for the case of $S=\{s\}$ and $m = 2s+1$, there exists at least one user who can decode $s+1$ messages. From this, the extension to the ``consecutive complete--$S$'' PICOD follows.

\end{abstract}

\section{Introduction}  
\label{sec:introduction}

In the Index Coding (IC) problem, there is one sender/transmitter with $m$ independent messages to be delivered to $n$ clients/users through an error-free broadcast link.
Each user has some message as side information (i.e., a subset of the message set) available to it 
and needs to reliably decode some messages that are not in its side information set.
In the IC problem, the desired messages for each user are pre-determined
and one asks what is the minimum number of transmissions (i.e., minimum code length)
such that every user is able to decode its desired messages successfully~\cite{index_coding_with_sideinfo}. 
The general IC is open.
When one restricts attention to linear codes, 
finding the minimal code length is equivalent to the so-called minrank problem, which is NP-complete~\cite{index_coding_with_sideinfo}.
Since IC is equivalent to the general network coding problem, it is known that linear schemes are not sufficient in general~\cite{linear_suboptimality_IC}.

A relaxed version of the IC, known as Pliable Index CODing (PICOD) has recently attracted attention~\cite{BrahmaFragouli-IT1115-7254174}.
The difference between PICOD and IC is that for PICOD the desired messages by the users are not pre-determined, that is, each user can choose to decode {\it any message 
not in its side information set}. The goal in PICOD is to find the desired message assignment that minimizes the code length.
The freedom of choosing the desired message to decode results in significant reduction in PICOD number of transmission/code length compared to the classical IC problem with the same number of message, number of users, and message side information sets~\cite{BrahmaFragouli-IT1115-7254174}.

\paragraph*{\bf Past Work on PICOD}
Known achievability schemes are based on linear codes only, and
very few converse results are available.
For the oblivious/complete--$\{s\}$ PICOD problem with $m$ messages~\cite{BrahmaFragouli-IT1115-7254174} (i.e., there are $n = \binom{m}{s}$ users each with a distinct side information set of cardinality $s\in[0:m-1]$) the optimal code length under the restriction that the sender can only employ linear codes is known to be $\min\{m-s,1+s\}$~\cite[Th.~9]{BrahmaFragouli-IT1115-7254174}.
For converse results, our work~\cite{itw_2017} provided the first information theoretic converse for some classes of 
complete--$S$ PICOD problems (formally defined later) 
and for PICOD problems where the topology hypergraph is an circular-arc hypergraph; our converse results show that linear codes 
are information theoretic optimal for those cases. 
The objective of this paper is to prove information theoretic converse results for more general classes of PICOD than in~\cite{itw_2017}.

\paragraph*{\bf Contributions}
In this paper we derive a tight information theoretic converse for some complete--$S$ PICOD settings not covered in~\cite{itw_2017}. A complete--$S$ PICOD is a system with $m$ messages where all side information sets/users with size indexed by $S$ are present, where $S \subseteq[0:m-1]$. 
In particular, we focus here on the ``consecutive case'' where 
$S$ satisfies $S=[\smin:\smax]$ for some 
$0\leq \smin\leq \smax \leq m-1$.

%
As in~\cite{itw_2017}, our converse is based on showing the existence of at least one ``special user'' who can decode a certain number of messages outside its side information set; the stumbling block is how to find such a ``special user.''
Compared to~\cite{itw_2017}, here we approach the problem by using a novel combinatorial technique: instead of  {\it constructively finding} such a ``special user'' for each choices of desired messages, 
we show the existence of this ``special user'' regardless of the choice of desired messages.
This is accomplished by not simply focusing on the message a user desires to decode, but on all the messages that a user will eventually be able to decode.
This new technique greatly reduces the complexity of the proof compared to~\cite{itw_2017} 
and enables us to obtain a converse bound for a very general class of complete--$S$ PICOD. 
Our result shows that the general converse bound in~\cite[Prop.1]{itw_2017} is loose  for the ``consecutive complete--$S$'' PICOD, whereas the newly proposed bound is tight and achieved by linear codes.
The keystone of the proof is to show that, for the ``critical case'' of $S=\{s\}$ and $m = 2s+1$, there exists at least one user 
who can decode $s+1$ messages. From this, the extension to the ``consecutive complete--$S$'' PICOD follows  by enhancing the system to a ``critical case'' one. 

\paragraph*{\bf Paper Organization}
Section~\ref{sec:system_model} introduces the system model; 
Section~\ref{sec:result} presents our tight converse result;
Section~\ref{sec:conclusion} concludes the paper;
proofs are in Appendix.

\section{System Model}  
\label{sec:system_model}

\paragraph*{\bf Notation}
Throughout the paper we use 
capital letters to denote sets,
calligraphic letters for family of sets, 
and lower case letters for elements in a set. 
The cardinality of the set $A$ is denoted by $|A|$. 
For integers $a_1\leq a_2$ 
we let $[a_1:a_2] := \{a_1,a_1+1,\ldots,a_2\}$ 
and $[a_2]:=[1:a_2]$ for $a_2\geq 1$.
For a set $W$ and an index set $A$, we let $W_A:=\{w_a\in W : a\in A\}$. 

\paragraph*{\bf System Model}
In a PICOD system there is one server and $n$ users, 
$u_{1},u_{2},\ldots,u_{n}.$ 
The server is connected to all users with a rate-limited noiseless broadcast channel. 
There are $m$ independent and uniformly distributed binary messages of $k$ bits. 
The message set is denoted as $W := \left\{ w_{1},w_{2},\ldots,w_{m}\right\}$.
User $u_i, i\in[n],$ has a partial knowledge of the message set as its side information $A_i\subset [m]$. 
The collection $\mathcal{A} := \{A_1, \ldots, A_n\}$
is assumed globally known. 
The server broadcasts 
a codeword of length $\ell k$ bits, which is a function of the message set $W$ and the collection of the side information sets $\mathcal{A}$, i.e., 
	$x^{\ell k} = \mathsf{ENC}(W,\mathcal{A})$.	
Each user decodes based on the code $x^{\ell k}$ and its own side information set;
for user $j\in[n]$, the decoding function is 
	$\widehat{w}_{j} = \mathsf{DEC}_j(W_{A_j},x^{\ell k}) \in W$.
A code is said to be \emph{valid} if every user can successfully decode at least one message not in its side information,  i.e., for user $j\in[n]$ there exists an index $d_j \notin A_j$
such that for some $\epsilon\in(0,1)$ 
\begin{align*}
	\Pr[ \exists d_j \in [m] \setminus A_j : \widehat{w}_{j}\neq w_{d_j} 
	]  \leq \epsilon;
\end{align*}
message $w_{d_j}$ is referred to as the \emph{desired message} by user $u_j$. 
For a valid code, the choice of desired messages 
is the set of indices 
$D := \{d_1,d_2,\ldots\,d_n\}$ where $d_j\in[m] \backslash A_j, \forall j\in [n]$.
The goal is to find a valid code with minimum length,
 i.e., 
\begin{align*}
	\ell^*:= \min\{\ell : \text{$\exists$ valid code of length $\ell k$ bits, for some $k$}\}.	
\end{align*}

\section{Main Result}  
\label{sec:result} 

In this paper we focus on a subclass of PICOD problems.
The {\em complete--$S$ PICOD problem},
for a given set $S\subseteq[0:m-1]$ where $m$ is the number of messages, consists of $n=\sum_{s\in S}\binom{m}{s}$ users, where no two users have the same side information set. 
In other words, all possible users with distinct side information sets that are subsets of size $s$
of the $m$ messages, for all $s\in S$, are present.
The main result of this paper is the  following theorem.
\begin{thm}
\label{thm:complete_s_general}
For the {\it consecutive complete--$S$ PICOD problem}, where $S=[\smin:\smax]$ for integers $\smin,\smax$ such that $0\leq \smin \leq \smax\leq m-1$, the optimal code length is 
\begin{align}
\ell^*= \min\{m-\smin,\smax+1\},
\end{align}
which is achieved by linear codes.\hfill$\square$
\end{thm}

\begin{remark}\label{rem:picod(t)}
We conjecture that for the complete--$[\smin:\smax]$ PICOD$(t)$ problem with $m$ messages, 
for some $0\leq \smin \leq \smax\leq m-t$, 
where each user must decode at least $t$ messages that are not in its side information set,  the achievable code length $ \min\{m-\smin,\smax+t\}$ is optimal. Theorem~\ref{thm:complete_s_general} shows that the conjecture is true for the case $t=1$.\hfill$\square$
\end{remark}

The rest of this section contains the proof of Theorem~\ref{thm:complete_s_general} and is divided as follows:
Section~\ref{sub:achievability} contains the achievability argument; 
Sections~\ref{sub:DecodingChain} and~\ref{sub:MAIS} introduce the notions of ``Decoding Chain'' and
of ``Maximum Acyclic Induced Subgraph,'' respectively;
Section~\ref{sub:critical_case_and_intuition_of_converse_proof} gives the converse argument for the complete--$S$ PICOD with $S=\{s\}$ and $m=2s+1$ (referred to as the ``critical case'' as   
all other cases in the ``consecutive complete--$S$'' PICOD can be derived from it), and 
Section~\ref{sub:complete_s_cardinality_one} generalizes it to any $|S|=1$; finally
Section~\ref{sub:complete_s_consecutive} 
proves the case $S=[\smin:\smax]$. 

%

\subsection{Achievability}  
\label{sub:achievability}

We use two types of linear codes: 
\begin{enumerate}
	\item 
	Transmit $\left.\smax+t\right|_{t=1}$ messages, one by one. 
	With this, every user can decode at least $t=1$ message not in its side information set. 
	\item 
	Transmit $m-\smin$ linearly independent linear combinations of all messages, e.g., an MDS code that allows to recover from any $\smin$ erasures of $m$ symbols. 
	Since each user has at least $\smin$ messages in its side information, by receiving $m-\smin$ linear combinations, each user is able to decode all the messages not in its side information set.
\end{enumerate}
By using the code among the above two that has the shortest length, we have
$\ell^\star \leq \min\left.\{m-\smin,\smax+t\}\right|_{t=1}$.

\subsection{Converse Main Ingredient~1: Decoding Chain}
\label{sub:DecodingChain}

We will start the converse proof by showing that for the complete--$S$ PICOD with 
\begin{align}
\label{eq:critical}
S=\{s\}, \ m=2s+1 \ \text{(referred to as ``critical case'')}
\end{align} 
the optimal code length is $\ell^\star =s+1,$ which is equivalent to showing that there exists at least one user who can decode all the $s+1$ messages not in its side information set. To do so we need to introduce a couple of concepts (Decoding Chain and Maximum Acyclic Induced Subgraph) and their properties. We shall do this in this and the next subsection.

Consider a system where user $u_j$, who has side information $A_j$, desires message $d_j$. 
After decoding $w_{d_j}$, user $u_j$ knows messages $W_{A_j\cup\{d_j\}}$. 
Besides user $u_j$, there are $s$ other users whose side information sets are subsets of $A_j\cup\{d_j\}$. 
If any of these other users decode a message $w_{k}$ such that $k\notin A_j\cup\{d_j\}$, then user $u_j$ can decode message $w_k$ as well (because it has the same side information $A_k\subset A_j\cup\{d_j\}$ as $u_k$).
This reasoning can be repeated until 
 user $u_j$ can not longer mimic other users / decode extra messages.
Therefore, we have identified a ``decoding chain'' for user $u_j$. 

\begin{example}\label{ex:s=1,m=3}
Consider the complete--$\{1\}$ PICOD, i.e., $s=1$, $m=2s+1=3$, $n=\binom{m}{s}=3$ and $\ell^\star=s+1=2$. 
Say that
$u_1$ knows $A_1=\{1\}$ and desires $d_1=2$;
$u_2$ knows $A_2=\{2\}$ and desires $d_2=1$; and
$u_3$ knows $A_3=\{3\}$ and desires $d_3=1$.
By sending $w_1$, users $u_2$ and $u_3$ are satisfied;
by sending $w_2$, user $u_1$ is satisfied.
By the ``decoding chain'' argument, user $u_3$ is able to mimic $u_1$ (because he decodes the message that is the side information set of user $u_1$)
and therefore can also decode $w_2$; on the contrary,
users $u_2$ and $u_3$ will not be able to decode any more messages other than the desired one.
\hfill$\square$
\end{example}

As Example~\ref{ex:s=1,m=3} shows, 
for a specific user, there always is a choice of desired messages such that this user cannot decode any message other the desired one.
However, we shall prove that regardless of the choice of desired messages, there always exists a user who can decode $s+1$ messages.
Since there are $(s+1)^{\binom{2s+1}{s}}$ (doubly exponential in $s$) possible choices of desired messages, finding explicitly such a user for every case is intractable.
Therefore, for our converse, we shall show the existence of one such user. 
The key proof idea is as follows.

Instead of considering what message each user desires, we 
reason on the ``decoding chain'' for that user. 
For the ``critical case'' in~\eqref{eq:critical} we aim to show that there is a user who decodes $s+1$ messages (as in Example~\ref{ex:s=1,m=3}).
Arguing by contradiction, assume no user can decode $s+1$ messages,
this is, that every user can decode at least one but at most $s$ messages.
In other words, including the side information set, after receiving a valid code every user eventually know at least $s+1$ but at most $2s$ messages.
Let user $u_j$, with side information $A_j$, eventually decode the messages indexed by $B_j$.  
One can think of the set $C_j:=A_j\cup B_j$ as a ``block'' that ``covers''  $A_j$, 
by which we mean that
the set $C_j$ is a proper superset of $A_j$, 
user $u_j$ can mimic any users $u_k$ whose side information $A_k\subset C_j$, and $d_k\in C_j$ if $A_k\subset C_j$.
Therefore, for any subset of users we can find a collection $\mathcal{C}$ such that, for every side information $A_j$, there is a cover $ C_j\in \mathcal{C}$ such that $C_j\supset A_j$.

This ``block cover'' idea was inspired by the generalized Steiner system in combinatorial design~\cite{generalized_steiner_system}.
An $\mathcal{S}(s,*,m)$ Steiner system consists of blocks/sets that cover exactly once every subset of size $s$ 
from the ground set of size $m$. 
In a PICOD setting, we also have to cover all $s$-element subsets of $[m]$ (i.e., all users' side information sets), but our problem is not a generalized Steiner system because 
an $s$-element subset may be contained in more than one block as long as it is not an intersection of some collection of other blocks.
Therefore, our ``block cover'' is a relaxed generalized Steiner system.

Note that our argument by contradiction for the ``critical case'' in~\eqref{eq:critical} is equivalent to showing that a ``block cover'' with size at most $2s$ cannot exist.
Inspired by Steiner systems, our combinatorial proof shows 
the assumption that there is a choice of desired message such that $1\leq |B_j|\leq s, \forall j\in[n]$ leads to a contradiction, and thus
there must exist a user whose block cover has size $m=2s+1$.


\subsection{Converse Main Ingredient~2: Maximum Acyclic Induced Subgraph (MAIS)}
\label{sub:MAIS}


Recall that for a PICOD problem, each user decodes one message outside its side information set indexed by $D=\{d_1,d_2,\dots,d_n\}$. 
Once $D$ is fixed, PICOD reduces to a {\it multi-cast IC} problem (because a message may be desired by more than one user).
Similarly to the classic all-unicast IC problem, we can represent the side information sets and the desired message in a digraph~\cite{index_coding_with_sideinfo}.
Pick a subset $U\subseteq[n]$ of users who desire different messages and create a digraph $\mathsf{G}(U)$ 
as follows.
The vertices of $\mathsf{G}$ are denoted by $V(\mathsf{G})\subseteq W$ and are the desired messages by the users in $U$. 
A directed arc $(w_i,w_j)\in E(\mathsf{G})$ exists if and only if the user who desires $w_i$ has $w_j$ in its side information set. 
$\mathsf{G}$ is called acyclic if it does not contain a directed cycle.
The size of $\mathsf{G}$ is the number of its vertices $|V(\mathsf{G})|$. 
For PICOD, the 
MAIS is the acyclic induced subgraph on the digraph created by the choice of users that desire different messages such that no other choice of users produces an acyclic induced subgraph of larger size. 
Since \text{MAIS} depends on the desired message set $D$, we denote its size as $|\text{MAIS}(D)|$.

For PICOD, as for multi-cast IC, the size of MAIS is a converse bound on $\ell$~\cite{index_coding_with_sideinfo}, i.e., $\ell\geq |\text{MAIS}(D)|$. 
Finding \text{MAIS} in a digraph is an NP-hard problem~\cite{21np-hard-problems}.
Finding \text{MAIS} for the multi-cast IC problem is more difficult since one needs to check every possible choice of users with distinct desired messages. Since each choice of $D$ in PICOD corresponds to a multi-cast IC problem, and since in PICOD we must find the best $D$, finding \text{MAIS} for PICOD appears intractable.
Therefore, we shall not find the exact user for every given $D$ that has a desired property, i.e., decode a certain number of messages, but rather show that for every $D$ a user with the desired property exists.
%
Towards this goal, we have the following observations on \text{MAIS} 
for the ``critical case'' in~\eqref{eq:critical}  (proofs can be found in Appendix):
%
\begin{enumerate*} 

\item {[Claim~1]}
	\label{claim:mais=s+1_equal_desire_s+1}
	For a given $D$, $|{\text{MAIS}}(D)|=s+1$ if and only if there exists a user who can decode $s+1$ messages.	

\item {[Claim~2]}
	\label{claim:mais=s_counter}
	If there exists a $D$ such that $|\text{MAIS}(D)|<s+1$, there must exist a $D^\prime$ with $|\text{MAIS}(D^\prime)|=s$. 
\suspend{enumerate*}
Our argument by contradiction for the ``critical case'' in~\eqref{eq:critical}   is equivalent to showing that $|\text{MAIS}(D^\prime)|=s$ is impossible, i.e.,
we shall show that, given a valid code for such a $D^\prime$, 
there exists a user who can decode $s+1$ messages. 
For such a user, we can find a set of $s+1$ users who decode different messages which form an acyclic digraph. 
This contradicts to the condition that $|\text{MAIS}(D^\prime)|=s$, therefore $D^\prime$ does not exist;
by Claim~\ref{claim:mais=s_counter} $|\text{MAIS}(D)|\leq s$ is thus impossible and we must thus have $|\text{MAIS}(D)|\geq s+1$. Since there are at most $s+1$ messages not known by any given user, we must therefore have $|\text{MAIS}(D)| = s+1$.
%
Note: Claim~\ref{claim:mais=s+1_equal_desire_s+1} shows that $|\text{MAIS}(D)| = s+1$ is equivalent to the existence of a user $u_j$ with cover $C_j=[2s+1]$.



\subsection{Converse for the Critical Case in~\eqref{eq:critical}}
\label{sub:critical_case_and_intuition_of_converse_proof}

Towards proving Theorem~\ref{thm:complete_s_general} in full generality, 
we first focus on the ``critical case'' in~\eqref{eq:critical}.
We shall prove that for
 $S=\{s\}$ and $m=2s+1$, the optimal number of transmission is $\ell^*= s+1$,
 in particular, that
 there always exists a user who can decode $s+1$ messages,  which, by Claim~\ref{claim:mais=s+1_equal_desire_s+1}, is equivalent to $|\text{MAIS}(D)|=s+1$ for all $D$.
By contradiction, we assume that $|\text{MAIS}(D)|<s+1$ for some $D$, and thus by Claim~\ref{claim:mais=s_counter} there must exist a $D^\prime$ such that $|\text{MAIS}(D^\prime)|=s$.
The assumption that $|\text{MAIS}(D^\prime)|=s$ implies that one can find a set of  $s$ users, denoted by $V$, who desire different messages and with the strict partial order 
on $V$ given by: for distinct $i,j\in V$, if $i<j$ then $d_j\notin A_i$. 
Without loss of generality, let the desired messages by the users $V$ be $[s+2:2s+1]$. 
It is easy to see  (by the definition of \text{MAIS}) that 
with side information $[s+1]$, one is able to decode all the remaining messages in $[s+2:2s+1]$.

Consider the following $s+1$ users: for $i\in [s+1]$ user $u_i$ has side information $A_i = [s+1]\setminus\{i\}$. 
We have two cases.

{\it Case~a)} Assume that for some $k\in[s+1]$ we have $B_k\cap[s+1]=[s+1]\backslash A_k$ (where $B_k$ is the set of messages that user $u_k$ can decode and $A_k$ its side information). 
Since this user can know all messages $W_{[s+1]}$, it can decode all the remaining messages $W_{[s+2:2s+1]}$. Eventually this user decodes $s+1$ messages, $C_k=[2s+1]$.

{\it Case~b)} 
For every user $u_i, i\in[s+1],$
we have $B_i\subseteq [s+2:2s+1]$. 
We have the following claims 
(proofs can be found in Appendix):
\resume{enumerate*}

\item {[Claim~3]}
	\label{claim:intersection_not_s}
	For the setting in this Case~b,  for any $P\subseteq [s+1]$, we have  $|\cap_{i\in P}B_i|\neq |P|-1$.

\item {[Claim~4]}
\label{thm:exist_intersection_s}
	For $s+1$ arbitrary subsets $B_i$ 
	from a ground set of size $s$, there always exists a set $P\subseteq [s+1]$ such that $|\cap_{i\in P}B_i|=|P|-1$.
\end{enumerate*}
Since Claims~\ref{claim:intersection_not_s} and~\ref{thm:exist_intersection_s} contradict each other, we have that Case~b is impossible.
Case~a shows 
the existence of a user whose block cover is $[m]=[2s+1]$.
Overall, this shows that for all possible choices of $D$ one must have $|\text{MAIS}(D)|=s+1$, which implies $\ell^*\geq s+1$. This, with the achievability in Section~\ref{sub:achievability}, concludes the proof of Theorem~\ref{thm:complete_s_general} for the ``critical case'' in~\eqref{eq:critical}, i.e., $\ell^* = s+1$.

%

\subsection{Converse for the Complete--$S$ PICOD with $|S|=1$}  
\label{sub:complete_s_cardinality_one}

In Section~\ref{sub:critical_case_and_intuition_of_converse_proof} we proved Theorem~\ref{thm:complete_s_general} for the complete--$S$ PICOD with $S=\{s\}$ and $m=2s+1$. Here we extend it to the cases 
%
$m<2s+1$ and $m>2s+1$, thus exhausting all complete--$S$ PICODs with $|S|=1$.

\subsubsection{Complete--$\{s\}$ PICOD with $m<2s+1$: $\ell^*= m-s$}  
%
	Consider a complete--$\{s\}$ PICOD problem with $m<2s+1$ and an integer $\alpha \leq s$.
The $n=\binom{m}{s}$ users in the system can be split into two categories: 
users $u_i$ with $[\alpha] \subset A_i$, 
and the other users.
The users in the first category 
do not decode any message in $[\alpha]$ (as their are in their side information set);
these users together form a complete--$\{s-\alpha\}$ PICOD with $m-\alpha$ messages.
Since this complete--$\{s-\alpha\}$ PICOD is a subset of the original complete--$\{s\}$ PICOD, its optimal number of transmissions is a lower bound on the number of transmissions in the original system.
If we take $m-\alpha = 2(s-\alpha)+1 \Longleftrightarrow \alpha = 2s+1-m > 0$ then, by the result in Section~\ref{sub:critical_case_and_intuition_of_converse_proof}, the optimal number of transmissions for the complete--$\{s-\alpha\}$ PICOD with $m-\alpha$ messages is $(s-\alpha)+1=m-s$.

Therefore the original complete--$\{s\}$ PICOD 
requires at least $m-s$ transmissions, i.e., $\ell^*\geq m-s = \min\{m-s,s+1\}.$

\subsubsection{Complete--$\{s\}$ PICOD with $m>2s+1$: $\ell^*= s+1$}  
\label{ssub:proof_fo_case_large_m}
%
	The proof is by contradiction.
	Assume there exists a $D$ such that $|\text{MAIS}(D)|=s$ and, 
	without loss of generality, that the maximum acyclic induced subgraph is formed by users with desired messages $[s]$.
	Specifically, we have users $u_i,i\in[s]$ such that $d_i=i$ and $d_j\notin A_i$ for any $j,i\in [s], j>i$ (by the definition of \text{MAIS} and its induced partial order).

	Let $U^\prime$ index the users whose side information is a subset of $[s+1:m]$, i.e., $i\in U^\prime$ if $A_i\subset [s+1:m]$.
	Apparently $1\in U^\prime$. We distinguish two cases.
	
{\it Case~a)}
	If there is a user $u_t\in U^\prime$ with desired message $d_t\in [s+1:m]$, we have $d_j\notin A_t$ for all $j\in[s]$. Thus users $u_t,u_1,u_2,\dots,u_s$ form an acyclic induced subgraph of length $s+1$. This contradicts to the assumption that $|\text{MAIS}(D)|=s$.

{\it Case~b)}
	For all $t\in U^\prime$ we have $d_t\in [s]$.
	By reasoning as in Section~\ref{sub:critical_case_and_intuition_of_converse_proof}, we can show that there exists a user who can decode $s+1$ messages. This again contradicts the assumption that $|\text{MAIS}(D)|=s$. 

	By combining the two above cases, we conclude that $|\text{MAIS}(D)| > s$.
	By Claims~\ref{claim:mais=s+1_equal_desire_s+1} and~\ref{claim:mais=s_counter} we thus have $\ell^*\geq s+1$.


\subsubsection{Complete--$\{s\}$ PICOD}
We showed that for the complete--$\{s\}$ PICOD we have $\ell^*\geq \min\{m-s,s+1\}$.
This, with the achievability in Section~\ref{sub:achievability}, concludes the proof of Theorem~\ref{thm:complete_s_general} for the case $|S|=1$, i.e., $\ell^* = \min\{m-s,s+1\}$.

\subsection{Converse for the Complete--$[\smin:\smax]$ PICOD}  
\label{sub:complete_s_consecutive}

With the result in Section~\ref{sub:complete_s_cardinality_one},
we are ready to prove Theorem~\ref{thm:complete_s_general} in full generality.
We consider three cases.

\subsubsection{Case $\smax \leq \ceil{m/2}-1$: $\ell^*=\smax+1$} 
\label{prop:max<m/2}

Drop all the users except those with side information of size $\smax$, thereby obtaining a compete-$\{\smax\}$ PICOD with $m$ messages; for this system, the optimal number of transmissions is $\min\{m-\smax,\smax+1\} = \smax+1$ (because  $\smax+1 \leq \ceil{m/2}$ in this case), which is a lower bound on the number of transmissions in the original systems. By our first type of achievability in Section~\ref{sub:achievability}, we have $\ell^*=\smax+1$.

\subsubsection{Case $\smin \geq \floor{m/2}$: $\ell^*=m-\smin$}
\label{prop:min>m/2}
As for the case in Section~\ref{prop:max<m/2}, drop all the users except those with side information of size $\smin$, thereby obtaining a compete-$\{\smin\}$ PICOD with $m$ messages and optimal number of transmissions is $\min\{m-\smin,\smin+1\} = m-\smin$ (because  $\smin \geq \floor{m/2}$ in this case). This lower bound on the number of transmissions in the original systems is attained by our second type of achievability in Section~\ref{sub:achievability}.

\subsubsection{Case $\smin \leq \ceil{m/2}-1 \leq \floor{m/2} \leq \smax$} 
\label{prop:min<m/2<max}
 	Define $\delta := \min\{\smax-\floor{m/2}, \ceil{m/2}-1-\smin\}$, drop all users except those with side information of size $s\in [\ceil{m/2}-1-\delta: \floor{m/2}+\delta]$, thereby obtaining a complete--$[\ceil{m/2}-1-\delta: \floor{m/2}+\delta]$ PICOD with $m$ messages.
 	Create dummy messages $W_{[m+1: m^\prime]}$, where $m^\prime=m+2\delta+\floor{m/2}-\ceil{m/2}+1$. Dummy messages will not be desired by any user.
To every user who was not dropped and has size information set of size $s\in [\ceil{m/2}-1-\delta: \floor{m/2}+\delta]$ 
give, as extra side information, an $(\floor{m/2}+\delta-s)$-subset of $[m+1: m^\prime]$; 
each such user 
generates $2\delta+\floor{m/2}-\ceil{m/2}+1 \choose \floor{m/2}+\delta-s$ new users.
 	This procedure gives a complete--$\{\floor{m/2}+\delta\}$ PICOD with $m^\prime$ messages, whose optimal number of transmissions is 
	\begin{align*}
	&\min\{\floor{m/2}+\delta+1, m^\prime-(\floor{m/2}+\delta)\} 
  \\&=\min\{\smax-\floor{m/2}+\floor{m/2}+1, \ceil{m/2}-\smin+\floor{m/2}\}
  \\&=\min\{\smax+1, m-\smin\} =: \ell^\prime.
	\end{align*} 
	Although the new system contains more users, any valid code for the original system works for the new one.
	Therefore $\ell^\prime$ is a lower bound on the optimal number of transmissions for the original system. This lower bound can be attained by the scheme described in Section~\ref{sub:achievability}.

\section{Conclusion}
\label{sec:conclusion}

This paper proved that simple linear codes are information theoretically optimal for the ``consecutive complete--$S$'' PICOD problem.  The main contribution is a novel way to deal with the optimization over the different choices of users' desired message.
The new proof technique is inspired by combinatorial design~\cite{generalized_steiner_system,a_course_in_combinatorics}.
It relates the problem of finding a user with certain properties to the existence of a  ``block cover'' with certain properties, where a ``block'' includes all messages that a user can eventually decode; this in turns is related to the the size of \text{MAIS} for the resulting multi-cast index coding problem.
This combinatorial approach overcomes a limitation of deriving general converse results in past work~\cite{itw_2017}  and cane extend beyond the ``consecutive complete--$S$'' PICOD case.



\appendix



\subsection{Proof of Claim~\ref{claim:mais=s+1_equal_desire_s+1}}
	If $|\text{MAIS}(D)|=s+1$, there are $s+1$ users who desire different messages. These users form an acyclic graph. An acyclic graph has an order, where the first user has side information that contains all messages not desired by these $s+1$ users. The first user, by decoding its desired message, can mimic the second user and then decode the desired message of the second user. This process goes on. Eventually the first user can mimic all the rest $s$ users and decode $s+1$ messages. 

	Conversely, if there exists a user who can decode $s+1$ messages, then $|\text{MAIS}(D)|=s+1$. For the user who can decode $s+1$ messages, it first decodes its desired message and then decodes other $s$ messages by mimicking other users. 
	These $s+1$ users form an acyclic subgraph of size $s+1$.

\subsection{Proof of Claim~\ref{claim:mais=s_counter}}
	Recall that $D=\{d_1,d_2,\ldots,n\}$ for $n=\binom{m}{s}$, are the indices of the desired messages by all users, with $d_j\in [m]\setminus A_j$ for all $j\in[n]$.
	Let there be an order of the $m$ messages, starting from $1$ to $m$. 
	Let $D_1=\{d_{11},\dots,d_{1n}\}$ where $d_{1j}$ is the smallest index that is not in $A_j$. $d_{1j}\in [s+1]$ for all $j\in[n]$, only the first $s+1$ messages are desired.
	Under $D_1$ the original PICOD comes a complete--$[0:s]$ PICOD with $m=s+1$, therefore $|\mais(D_1)|=s+1$.
	Now let us assume there is $D_k$ with $|\mais(D_k)|\leq s$.
	$D_k$ can be obtained from $D_1$ from the following steps:
	\begin{enumerate}
		\item If $d_{11}\neq d_{1k}$, change $d_{11}$ to the next message in the order that is not in the side information $A_1$. Name it $D_2$. Then compare $d_{12}$ and $D_{1k}$. Repeat until we have $d_{1j}=d_{1k}$.
		\item Move to the next entry. Repeat the same steps until it is the same to $d_{2k}$.
		\item Iterate until the last entry. 
	\end{enumerate}
	By these steps we create an order of desired message $D_1,D_2,\dots,D_k$.
	The adjacent choices $D_i,D_{i+1}, i\in[k-1]$ differ in only one entry in this order, i.e., all users but one desire the same messages.

	Recall that $|\text{MAIS}(D_i)|$ is the size of maximum induced acyclic subgraph by choosing users in $U^*_i$.
	From $D_i$ to $D_{i+1}$, in the digraph representation, only one vertex can change. 
	As a result, for any induced acyclic subgraph, only one vertex can change.
	The size of any induced acyclic subgraph is changed by at most 1.
	Since MAIS bound is essentially the size of some induced acyclic subgraph, we have $|\mais(D_{i+1})|\in [\mais(D_i)-1:\mais(D_i)+1]$, i.e., in the order
	the MAIS bounds of two adjacent choice of desired messages differ by at most one.
	We have $|\mais(D_1)|=s+1$ and $|\mais(D_k)|\leq s$.
	This shows that there exists $D^\prime$ such that $|\mais(D^\prime)|=s$.
	

\subsection{Proof of Claim~\ref{claim:intersection_not_s}}
	We assume that $B_i\subseteq [s+2:2s+1]$. 
	Note $B_i$ is the set of indices of the messages decoded by user $u_i$;
	by the ``decoding chain,'' for any user $u_k$ with $A_k\subset C_i=A_i\cup B_i$, we have $d_k\in C_i$.
	By definition of ``decoding chain,'' we have $|\cap_{i\in P}C_i|\neq s$ for any $P\subseteq [s+1]$.
	This is so because if $|\cap_{i\in P}C_i|= s$, we have $\cap_{i\in P}C_i= A_k$ for some $k\in[n]$. 
	Then $d_k\in C_i,\forall i\in P$ since all users 
	indexed by $P$ can mimic it.
	However, $d_k\notin A_k=\cap_{i\in P}C_i$ as $u_k$ needs to decode the message outside its side information.
	Therefore, $\exists i\in P$ such that $d_k\notin C_i$.
	We have a contradiction.
	Therefore $|\cap_{i\in P}C_i|\neq s$ for all $P\subseteq [s+1]$.
	Note that $|\cap_{i\in P}A_i|= s+1-|P|$ and $A_i\cap B_i = \emptyset$, 
	thus we have $|\cap_{i\in P}B_i|\neq |P|-1$.

 \subsection{Proof of Claim~\ref{thm:exist_intersection_s}}
 For the sake of space we give a brief outline of the proof, which is based on the following lemma.
 \begin{lemma}
	\label{lem:cross_lemma}
	For a $n\times m$ binary matrix with no all zero rows, there exists a pair $(i,j)$ such that entry $(i,j)$ is 1 and $\frac{\text{\# of $1$s in $i$th row} }{\text{\# of $1$s in $j$th column}}\geq \frac{n}{m}$.
\end{lemma}

	We proof Claim~\ref{thm:exist_intersection_s} by induction on $s$.

    When $|B_i|=0$ for some $i$, take $P=\{i\}$, we have $|\cap_{i\in P}B_i|=0=|P|-1$. 
    Therefore we only need to consider the case where all $B_i$ are non-empty.
    
    For the initial case $s=1$ the statement is true. 
    It can be seen since $B_1=B_2=\{1\}$. Take $P=[2]$ we have $|\cap_{i\in[2]}B_i|=1=2-1$. 

    Assume the statement is true for all $s\leq t-1$. 
    By Lemma~\ref{lem:cross_lemma} we can construct a $P\subseteq [s+1]$ such that $|\cap_{i\in P}B_i|=|P|-1$ for $s=t$.
    
    Therefore exists $P\in[s+1]$ such that $|\cap_{i\in P}B_i|=|P|-1$ for all positive integer $s$.
    

\bibliographystyle{IEEEtranS}
\bibliography{refs}

\IEEEtriggeratref{3}

\end{document}